\RequirePackage{fix-cm}
\documentclass[smallextended]{svjour3}       
\smartqed  
\usepackage{graphicx}
\usepackage{epsf}

\journalname{J Supercond Nov Magn}
\begin{document}

\title{Magnetically ordered state at correlated oxide interfaces: the role of random oxygen defects 
}

\titlerunning{Magnetic ordered state in heterostructures of titanates}        

\author{N. Pavlenko  \and
        T.~Kopp 
}


\institute{N. Pavlenko \at
              Experimental Physics VI and Theoretical Physics III, EKM, Institute of Physics, University of Augsburg, Universit\"atstr.1, 86135 Augsburg, Germany \\
              Tel.: +49-821-5983664\\
              Fax: +49-821-5983652\\
              \email{pavlenko@mailaps.org}\\           
             \emph{On leave from} the Institute for Condensed Matter Physics, NAS, 79011 Lviv, Ukraine \\ 
            \and
           T. Kopp \at
           Experimental Physics VI, EKM, Institute of Physics, University of Augsburg, Universit\"atstr.1, 86135 Augsburg, Germany   
}

\date{Received: April 29, 2012 / Accepted: April 29, 2012}

\maketitle 

\begin{abstract}
Using an effective one-band Hubbard model with disorder, we consider magnetic states of the correlated oxide interfaces, where effective hole self-doping and a magnetially ordered state emerge due to electronic and ionic reconstructions. By employing the coherent potential approximation, we analyze the effect of random oxygen vacancies on the two-dimensional magnetism. We find that the random vacancies enhance the ferromagnetically ordered state and stabilize a robust magnetization above a critical vacancy concentration of about $c=0.1$. In the strong-correlated regime, we also obtain a nonmonotonic increase of the magnetization upon an increase of vacancy concentration and a substantial increase of the magnetic moments, which can be realized at oxygen reduced high-$T_c$ cuprate interfaces.

\keywords{Oxide interfaces \and Titanates \and Magnetism \and Interface reconstruction}
\end{abstract}

\section{Introduction}
\label{intro}
In complex thin-film transition-metal oxide heterostructures, interface phenomena can substantially affect the electronic properties and lead new electronic ordered states not observed in the bulk constituents \cite{ohtomo1,pavlenko_kopp,pavlenko_kopp2,pavlenko_schwabl2,jackeli,pavlenko_sawatzky,pavlenko_orb}. Prominent examples are the titanate
interfaces in the heterostructures LaAlO$_3$/SrTiO$_3$, LaTiO$_3$/SrTiO$_3$ and similar compounds \cite{ohtomo,thiel}. These structures have recently
attracted much attention which is focused on the discovered two-dimensional electron liquid state stabilized by the interface electronic reconstruction. 
The magnetoresistance and torque magnetometry measurements \cite{brinkman,dikin,li} demonstrate a hysteretic behavior in the 
field-dependences of magnetoresistance and superparamagnetic behavior which suggests the existence of ferromagnetic puddles 
in the samples of LaAlO$_3$ (LAO) grown on SrTiO$_3$ (STO).

The observation of the magnetism at the LAO/STO interface triggered 
an intense exploration of the role of impurities in the formation of the magnetically ordered state at these interfaces \cite{ohtomo,brinkman,kalabukhov,dikin,li,ariando,pavlenko,pavlenko2,pavlenko3}.
In particular, recent first-principle studies demonstrate a vacancy-related magnetic exchange splitting of $3d$ states of interface Ti atoms~\cite{pavlenko2,pavlenko3}. In this case, the conducting electrons, which emerge due to the charge compensation of the polar discontinuity occupy $d_{xy}$ bands, whereas the vacancy-released electrons occupy $e_g$ states shifted below the Fermi level due to the interface orbital reconstruction ~\cite{pavlenko2,pavlenko3}.     
STM, cathode luminescence studies and conductivity measurements provide strong support for
the existence of the oxygen vacancies in STO layers of LAO/STO heterostructures \cite{muller,kalabukhov}.

As the DFT calculations are unable to access a realistic random low-concentration distribution of impurities,
we consider here an effective two-dimensional one-band Hubbard model on a square lattice with disordered random vacancies. In this model, each vacancy introduces an exchange splitting of the local $3d_{xy}$ state of neighbouring Ti atoms, in this way stabilizing a ferromagnetic order through the electronic transfer term.   

\section{One-orbital model of random oxygen vacancies}

\begin{figure}[tbp]
\epsfxsize=5.0cm {\epsfclipon\epsffile{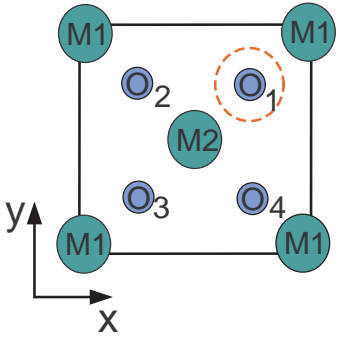}}
\caption{Scheme of a doubled unit cell on the MO$_2$-plane (M=Ti) of a SrTiO$_3$-layer 
}
\label{fig1}     
\end{figure}

The two-dimensional electronic liquid at the titanate interfaces is described by an effective one-band Hubbard model on a lattice with $N$ sites which corresponds to the interface TiO$_2$ layer. Each site $i$ identifies a doubled $\sqrt{2}\times \sqrt{2}$ TiO$_2$ unit cell with two Ti atoms $j=1,2$ and four nearest neighbouring oxygen atoms $l=1,\ldots4$ (see Figure~\ref{fig1}), where the cell doubling is introduced for the studies of magnetically ordered states on two Ti sublattices. In this configuration, the oxygen vacancy corresponds to the elimination of one of the oxygens in the Ti$_2$O$_4$-plaquette. 
The local disorder induced by an oxygen vacancy on the lattice site ($i,l$) is introduced through the local random fields $h_{i\sigma,l}=h_\sigma$, which shift the electronic $3d_{xy}$ states of the neighbouring Ti atoms: 
\begin{eqnarray}
H=&&(\varepsilon_d-\mu)\sum_{i=1\atop \sigma; j=1,2}^N d_{i\sigma,j}^\dag d_{i\sigma,j}+U\sum_{i\atop j=1,2} n_{i\uparrow,j} n_{i\downarrow,j} +T_{dd} \nonumber\\
&& +\sum_{i\sigma\atop l=1,\ldots 4} h_\sigma(1-x_{il})d_{i\sigma,2}^\dag d_{i\sigma,2}\\
&& +\sum_{\langle ii'\rangle\sigma \atop l=1,\ldots 4} h_\sigma(1-x_{i'l})d_{i\sigma,1}^\dag d_{i\sigma,1}, \nonumber\\
&& T_{dd}=t \left (\sum_{i ;\sigma} d_{i\sigma,1}^\dag d_{i\sigma,2}+\sum_{\langle ii'\rangle ;\sigma} d_{i\sigma,1}^\dag d_{i'\sigma,2}+h.c. \right).
\end{eqnarray}
Here $d_{i\sigma,j}^{\dag}$ are electron creation operators and $n_{i\sigma,j}=d_{i\sigma,j}^\dag d_{i\sigma,j}$ are the occupation numbers for the self-doped $3d_{xy}$ electrons with the local energy $\varepsilon_d$ and chemical potential $\mu$; $U$ is the local Hubbard repulsion and $t$ is the effective indirect $d-d$ electronic transfer energy. The binary discrete random variable $x_{il}=\{0,1\}$ is zero if the oxygen atom is absent in the oxygen position ($i$,$l$) of the unit cell $i$. The two last terms in $H$ describe the magnetic splitting $h_\uparrow =-h_\downarrow=-h$ of the local electronic states of Ti due to an oxygen vacancy in each of four possible neighbouring ($i,l$) positions \cite{pavlenko2,pavlenko3}. For Ti$_2$, the nearest neighbouring oxygen atoms belong also to different unit cells with the coordinates $\vec{R}_{i'}=\{\vec{R}_{i},\vec{R}_{i}-\vec{a}_x,\vec{R}_{i}+\vec{a}_y,\vec{R}_{i}-\vec{a}_x+\vec{a_y}\}$.   

In the ferromagnetic state, the local magnetic moment $m=m_j=\langle n_{i\uparrow,j}\rangle-\langle n_{i\downarrow,j} \rangle$ is defined through the average orbital occupancies of the majority versus the minority local spin states which can be expressed via the thermodynamic averages of the corresponding double-time one-particle Green functions $G_{i\sigma; i'\sigma'}^{jj'}(t-t')=-i\Theta(t-t')\langle [d_{i\sigma,j}(t),d_{i'\sigma,j'}^{\dag}(t')]\rangle$ \cite{zubarev}. Here $d_{i\sigma,j}(t)$ and $d_{i'\sigma,j'}^\dag(t')$ are the Heisenberg representations of the operators $d_{i\sigma,j}$ and $d_{i'\sigma,j'}^\dag$. To calculate the average electronic orbital occupancies for the Hamiltonian with the random configurational variables $x_{il}$, we employ the coherent potential approximation \cite{elliott,esterling} which allows to express the configurationally averaged Green functions $\langle G_{i\sigma;i'\sigma'}^{jj'}(t-t') \rangle_c$ through the effective medium Green functions $R_{i\sigma; i'\sigma'}^{jj'}(t-t')$: 
\begin{eqnarray}
G_{i\sigma; i'\sigma'}^{jj'}=R_{i\sigma; i'\sigma'}^{jj'}+\sum_{l=1\atop g=1,2}^N 
R_{i\sigma; l\sigma'}^{jg} T_g^l R_{l\sigma; i'\sigma'}^{gj'},
\end{eqnarray}
where the $T$ matrix $\{ T_g^l\}$ describes the local scattering on the defect potential.
In the $(\vec{k},\omega)$-space, the effective medium sublattice Green functions have the following form: 

\begin{eqnarray}\label{r}
R_{\sigma;\sigma}^{jj}(\vec{k},\omega)=R_{\sigma}^{jj}(\vec{k},\omega)=\frac{1}{2\pi D^\sigma}\left( \frac{x_1^\sigma-\varepsilon_{j}^\sigma}{\omega-x_1^\sigma}-\frac{x_2^\sigma-\varepsilon_{j}^\sigma}{\omega-x_2^\sigma} \right),
\end{eqnarray}
where $x_{1;2}^\sigma=(\varepsilon_1^\sigma+\varepsilon_2^\sigma)/{2}\pm {D^\sigma}/{2}$, $\varepsilon_{1;2}^\sigma=\varepsilon_d-\mu+U\langle n_{\bar{\sigma};1/2}\rangle+\Sigma_{1;2}^\sigma$, and $D^\sigma=\sqrt{(\varepsilon_1^\sigma-\varepsilon_2^\sigma)^2+4t^2|z_{\vec{k}}|^2}$ with $z_{\vec{k}}=1+\exp(ik_xa_x)+\exp(-ik_ya_y)+\exp(i\vec{k}(\vec{a}_x-\vec{a}_y))$ and $\langle n_{i\sigma;1/2}\rangle=\langle n_{\sigma;1/2}\rangle$. The effective self-energies $\Sigma_j$ should be determined from the equality of the effective-medium propagators and the corresponding configurationally averaged Green functions 
\begin{eqnarray}
R_{i\sigma;i'\sigma'}^{jj'}(t-t')=\langle G_{i\sigma;i'\sigma'}^{jj'}(t-t') \rangle_c,
\end{eqnarray}
which is equivalent to the condition $\langle T_g^l\rangle_c=0$ in the single-site approximation, and in our case leads to the following equations for the determination of $\Sigma_j$:
\begin{eqnarray}\label{t}
\langle V_j^{i\sigma}(1-g_j^\sigma V_j^{i\sigma})^{-1}\rangle_c=0; \quad (j=1,2),
\end{eqnarray}
where $V_1^{i\sigma}=h_\sigma (4-x_{i1}-x_{i-a_x,2}-x_{i+a_y,4}-x_{i-a_x+a_y,3})-\Sigma_1^\sigma$
and $V_2^{i\sigma}=h_\sigma (4-\sum_l x_{il})-\Sigma_2^\sigma$ are the random deviations from the effective medium self energies due to the local disorder. The local Green functions
\begin{eqnarray}\label{g0}
(g_j^\sigma)^{-1}=(G_{0j}^\sigma)^{-1}-\Sigma_j^\sigma
\end{eqnarray}
are determined from the bare Green functions for the stoichiometric lattice:
\begin{eqnarray}\label{gg0}
(G_{0j}^\sigma)^{-1}=\omega-\varepsilon_d+\mu-U\langle n_{\bar{\sigma};j}\rangle
\end{eqnarray}
which are calculated after the mean-field decoupling of the local Hubbard term in $H$. We note that the mean-field approach allows to capture the main features of the ordered states in the correlated systems and is widely used to study orbital physics at correlated interfaces \cite{jackeli,hirsch}.
  
In the ferromagnetically ordered state, we have 
$\Sigma_1^\sigma=\Sigma_2^\sigma$,
$G_{01}^\sigma=G_{02}^\sigma$, and the problem is reduced to the solution of the equations (\ref{t}) and the selfconsistent equations for the magnetization and chemical potential
\begin{eqnarray}\label{mn}
m=\langle n_{\sigma,j}\rangle - \langle n_{\bar{\sigma},j}\rangle\nonumber\\
n=\langle n_{\sigma,j}\rangle + \langle n_{\bar{\sigma},j}\rangle,
\end{eqnarray}
which should be considered for a given electron concentration $n$. The thermodynamically averaged occupancies $\langle n_{\sigma,j}\rangle$ in this case are expressed through the effective-medium Green functions $R_{\sigma;\sigma}^{jj}(\vec{k},\omega)$ and correspond to the configurationally averaged electronic occupation numbers. 

In the limit of small concentrations of the oxygen vacancies $c=\langle 1-x_{il}\rangle_c$, we can obtain the following expansion for $\Sigma_j^\sigma$ 
\begin{eqnarray}\label{sigma_c}
\Sigma_j^\sigma(\omega)=4h_\sigma \frac{c}{1-c}S_j(\omega)+O(c^2),
\end{eqnarray}
where $S_j(\omega)=1+{h_\sigma}/{(\omega-\varepsilon_d+\mu-U\langle n_{\bar{\sigma};j}\rangle-h_\sigma)}$. From (\ref{sigma_c}) we see that $\Sigma_j^\sigma \rightarrow 0$ for vanishing vacancy concentration $c \rightarrow 0$. 

\section{Results and discussion}

The numerical solutions of the equations (\ref{mn}) for different vacancy concentrations $c$ have been analyzed in the low-temperature range $0.0001<k_BT/t<0.01$ for the electronic concentration $n$ in the range between 0.1 and 0.4 which corresponds to the concentrations $6\cdot 10^{13}$~cm$^{-2}$--$2\cdot 10^{14}$~cm$^{-2}$ of the interface-doped polar charge in the TiO$_2$ layers measured in Hall effect experiments \cite{thiel,ohtomo} and estimated from ab-initio calculations \cite{pavlenko,pavlenko2,pavlenko3}. 

\begin{figure}[tbp]
\epsfxsize=8.0cm {\epsfclipon\epsffile{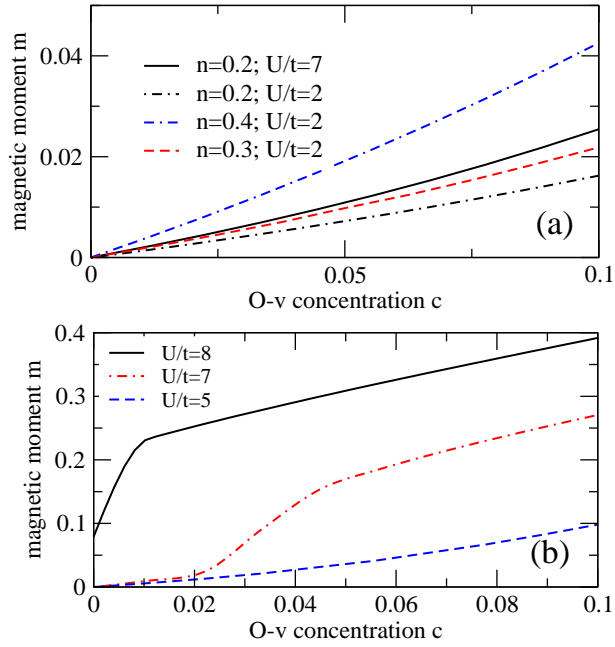}}
\caption{Local magnetic moment $m$ versus O-vacancy concentration $c$ for different $U$ and electron concentrations $n$. Here $h/t=0.25$, $k_BT/t=0.01$. In plot(b), the electron concentration is fixed to $n=0.4$.}
\label{fig2}       
\end{figure}

In the regime of weak and intermediate Hubbard correlation energies $U/t \le 5$, we obtained a monotonic increase of the local magnetization for higher vacancy concentrations $c$, which can be observed in Fig.~\ref{fig2}(a). The value $U/t \approx 2-5$ corresponds to the correlation energies of Ti $3d$ electrons used in ab-initio DFT+U calculations in \cite{pavlenko,pavlenko2,pavlenko3}. Fig.~\ref{fig2}(a) shows that a robust magnetic state with magnetic moments larger than $0.05$~$\mu_B$ can be achieved only for the high vacancy concentration $c>0.1$, which corresponds to vacancy densities above a critical value $6\cdot 10^{13}$~cm$^{-2}$. This range of vacancy densities was indeed analyzed in the ab-initio calculations which explains a good agreement of the results obtained in  \cite{pavlenko,pavlenko2,pavlenko3} with the experimental measurements.   

Furthermore, the analysis of the magnetic moments calculated in the range of strong Hubbard correlations $U/t>6$ shows the existence of a transition from the weak-magnetization regime to the regime of strong magnetism, indicated in Fig.~\ref{fig2}(b). In the strong-correlation regime, large values of the local magnetic moments of the order $0.3$~$\mu_B$ are stabilized already for small $c$, which can be explained by the intrinsic property of the Hubbard model to stabilize the ferromagnetic state in the concentration range of approximately quarter-filling: 
the transition from the paramagnetic to the ferromagnetic state upon an increase of $U$ can be identified from the magnetic phase diagram for the two-dimensional Hubbard model \cite{hirsch}, where the concentration range $0.2<c<0.5$ corresponds to the electronic doping levels considered in this work. Considering the large values of the magnetic moments, the situation at the cuprate interfaces corresponds rather to the strong-correlation regime. One can expect that the strong-correlation regime can be realized, for example, at oxygen-reduced cuprate interfaces, which would  
provide a possibility for a strong enhancement of the ferromagnetic state by oxygen vacancies, a scenario which may lead to the formation of ferromagmetic regions coexisting with a superconducting background.

\section*{Summary}

We considered magnetic states of correlated oxide interfaces, where effective charge self-doping and magnetically ordered states emerge due to the electronic and ionic reconstructions. Employing the coherent potential approximation to the effective one-band Hubbard model with disorder, we analyzed the effect of random oxygen vacancies on the formation of two-dimensional magnetism. We find that the random vacancies enhance the ferromagnetically ordered state and stabilize a robust magnetization above a critical vacancy concentration of about $c=0.1$. In the strong-correlated regime, we also obtain a nonmomnotonic increase of the magnetization upon an increase of vacancy concentration and observe a substantial increase of the magnetic moments. This enhancement appears due to
the intrinsic property of strong electron correlations to stabilize a ferromagnetic state at electron doping levels $0.3-0.5$ electrons per unit cell, typical for the polar-doped interfaces \cite{ohtomo,thiel}. Although a mean-field evaluation might well overestimate the tendency towards a ferromagnetic state, the inhomogeneous (disordered) states considered here rather support ferromagnetic correlations: disorder reduces the kinetic energy and the impurities, being effectively magnetic, present a seed for at least short-range magnetism.  

\begin{acknowledgements}
This work was  supported by the DFG (TRR~80), A.~von~Humboldt Foundation 
and by the Ministry of Education and Science of Ukraine (Grant No.~0110U001091). 
Grants of computer time from the UNL Holland Computing 
Center and Leibniz-Rechenzentrum M\"unchen through the SuperMUC project pr58pi are gratefully acknowledged. 
We wish to acknowledge very useful discussions with J.~Mannhart, G.A.Sawatzky, E.Y.Tsymbal, I.~Stasyuk and K.~Moler.
\end{acknowledgements}



\end{document}